\documentclass[aps,prl,letterpaper,11pt,twoside,tightenlines,nofootinbib,showpacs,preprint,twocolumn]{revtex4}
\usepackage{graphicx}
\usepackage[sort&compress]{natbib}
\usepackage{latexsym}
\usepackage{epsfig}
\usepackage{amsmath}

\newcommand{\be}{\begin{equation}}

\newcommand{\ee}{\end{equation}}
\newcommand{\bea}{\begin{eqnarray}}
\newcommand{\eea}{\end{eqnarray}}
\newcommand{\bef}{\begin{figure}}
\newcommand{\eef}{\end{figure}}
\newcommand{\bce}{\begin{center}}
\newcommand{\ece}{\end{center}}

\def\lsim{\mathrel{\rlap{\lower4pt\hbox{\hskip1pt$\sim$}}
    \raise1pt\hbox{$<$}}}         
\def\gsim{\mathrel{\rlap{\lower4pt\hbox{\hskip1pt$\sim$}}
    \raise1pt\hbox{$>$}}}         
\begin{document}
\title{Test of Space-Time Non-Commutativity at the Future Circular Collider}
\author{P. Castorina$^{1,2}$}
\affiliation{
\mbox{${}^1$ Dipartimento di Fisica, Universit\`a di Catania, Via Santa Sofia 64,
I-95123 Catania, Italy.}\\
\mbox{${}^2$ INFN, Sezione di Catania, I-95123 Catania, Italy.}\\
}

\date{\today}
\begin{abstract}

The Future Circular Collider (FCC) is a crucial step forward to study  new Physics beyond the standard model and  to test fundamental aspects as space-time minimal length and Lorentz violations.
As an example, a possible enhancement of $e^+e^-$ pair production due to non-commutative effects, catalyzed by the huge magnetic field produced at the beginning of a heavy ion collison at FCC, is discussed.
In noncommutative electrodynamics a free photon in the magnetic background can produce a $e^+e^-$ pair. In particular for hard  photons with transverse energy $100-600$ GeV at the beginning of the collision and for a particular kinematical setting of the pair , i.e. large total transverse momentum in the reaction plane and invariant mass in the range $200-400$ MeV, the non-commutative contribution, evaluated with the present bound of the non-commuativity fundamental area , can be significant. Other, more exotic, possible signatures  of space-time non-commutativity are also considered.

\end{abstract}
 \pacs{24.10 Pa,11.38 Mh,05.07 Ca}
 \maketitle

\section{Introduction}

The feasibility and potential role for new discoveries of a $pp$ collider at 100 TeV, the future circular collider (FCC), is currently under investigation \cite{mich1,mich2,mich3}
with the mail goal to study supersymmetry, dark matter, new bosonic and fermionic resonances   and the formation and properties of quark -gluon plasma in heavy ion collisions at very high energy, $\sqrt{s_{NN}} \simeq 39$ TeV \cite{armesto}.

The FCC proposal is certainly fascinating and this letter is  a brief appendix to the previous studies \cite{mich1,mich2,mich3,armesto}
to recall that by FCC one can discover not only new phenomena in particle Physics but test some fundamental aspects as space-time non-commutativity \cite{nc1,nc2,nc3} and violations of Lorentz invariance.

For the specific case considered here, this possibility relies on the huge magnetic field produced in a non-central relativistic heavy ion collisions \cite{dima1,dima2,dima3,sko}.
Indeed it has been shown \cite{dima1,dima2,dima3,sko} that at RHIC energy is possible to reach  strong magnetic field, $B \simeq m_{\pi}^2$ GeV$^2$, at the beginning of a very peripheral nucleus-nucleus collision
and the generated magnetic field can be approximately written as
\be
B(t)= \frac{B_0}{(1+t/\tau)^{3/2}}
\ee
with
\be
\tau=\frac{b}{2 sinh Y}
\ee
and 
\be
B_0= \frac{8 Z \alpha_{EM} sinh  Y }{b^2}
\ee
where $b$ is the impact factor, $Y$ is the beam rapidity and $Z,\alpha_{EM} = 1/137$ describe the electromagnetic coupling. 

For Au-Au  ($Z=79$) at LHC ($\sqrt{s} = 5.6$ TeV) and at FCC ($\sqrt{s} = 39$ TeV) one has respectively $Y=9.38, 11.33$, $\tau= 8.35*10^{-4}, 1.2*10^{-4}$ fm and
$B_0= 10.7, 74.4$ GeV$^2$ for $b = 10$ fm. Therefore at the beginning of the collision at FCC there is a huge magnetic field, order of magnitudes larger than the magnetar ones,
with the time dependence in eq.(1).

Some phenomenological consequences of the initial magnetic field have been investigated in ref. \cite{dima1,dima2,dima3,volo,dima4,dima5}. For the present analysis, one has to recall that a strong magnetic background field
catalyzes the effects of a fundamental length, i.e. a space-time non-commutativity described, for example, by the commutation relations \cite{nc1,nc2,nc3}
\be
[x_\mu,x_\nu] = \Theta_{\mu,\nu}
\ee
where $\Theta_{\mu,\nu}$ is a set of fixed (anti-symmetric) parameters. To preserve the unitarity of field theories in non-commutative space-time one assumes that $\Theta_{0i} = 0$ for $i=1,2,3$ and therefore
the non-commutative parameters can be characterized by a vector $\vec \theta$ according to the relation $\Theta_{ij} = \epsilon_{ijk} \theta_k$.

It should be clarified that space-time non-commutativity can be implemented with algebraic structures different from eq.(4), preserving for example the the Lorentz algebra through the introduction of dilated
generators \cite{bernardini1}, and that non-commutativity is not limited to space-time coordinates but can be studied in full phase space (see \cite{bernardini2,bernardini3} for recent analyses).

By considering the canonical formulation of non-commutativity in eq.(4), the corresponding non-commutative electrodynamics (NCED)  has been studied in ref. \cite{roman} and the most interesting aspect  is  the deformation of the dispersion relation. In particular, in NCED one can exactly show \cite{roman} that plane waves exist and, while those propagating along the direction of a background magnetic field $\vec{B}$ still travel at the usual speed of light, those which propagate transversely to $\vec{B}$ have a modified dispersion relation between energy, $\omega$ and momentum, $k$, given by 
\begin{equation}
 \omega = k (1 - \vec{\theta}_T \cdot \vec{B}_T).
\end{equation}
Many phenomenological consequences of NCED have been studied  \cite{noi1,noi2,noi3,noi4} on the basis of the modified lagrangian
\be
\begin{split}
\hat{I} =  - \frac{1}{4} \int d^4 x \; [F^{\mu \nu} F_{\mu \nu} 
 -  \frac{1}{2} \theta^{\alpha \beta} F_{\alpha \beta} F^{\mu \nu}F_{\mu \nu} \\ 
 + 2 \theta^{\alpha \beta} F_{\alpha \mu} F_{\beta \nu} F^{\mu \nu}]
\end{split}	
\ee
with $F_{\mu \nu} = \partial_{\mu} A_\nu - \partial_{\nu} A_\mu$, and $A_\mu$ the usual Abelian gauge field, obtained by the Seiberg-Witten \cite{sw} map to first-order in $\theta$. 
However due to tight bound on $\theta \le O[(1/10 TeV)^2]$ \cite{solito,orfeo1,orfeo2}, the non-commutative effects  are essentially not detectable unless a huge background fiels is present.

Indeed, by previous equation, a free non -commutative photon with energy $E_\gamma$
can produce a  $e^+e^-$ pair \cite{noi5} if the magnetic background is such that
\begin{equation}
 - 2 E_\gamma ^2 ( \vec \theta _T \cdot \vec B_T) =( p_+ + p_-)^2 > 4 m_e^2,
\end{equation}
where  $p_\gamma, p_+$ and $p_-$ are the four momenta of $\gamma$, of  $e^+$ and of $e^-$,
respectively.
\begin{figure}
{{\epsfig{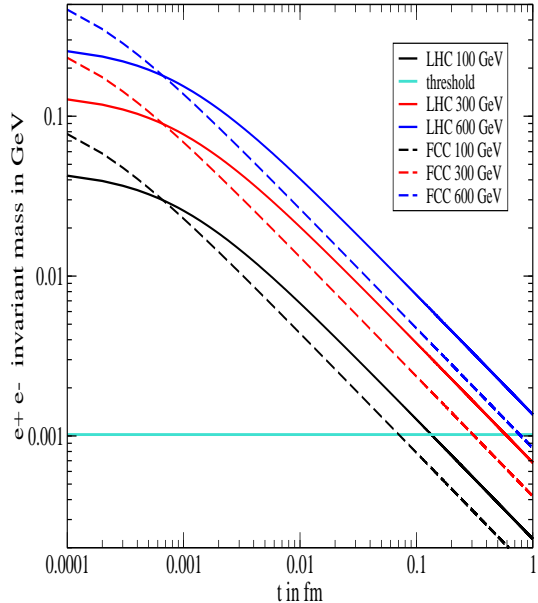}}
\caption{Time from the beginning of the collision when the non-commutative effects disappear, for diffferent $\gamma$ energy and beam energy. The most favourable case is considered, i.e.
$-\vec \theta_T \cdot \vec B_T = + \theta B$.
}}
\end{figure}

By assuming the most favorable condition $-(\vec \theta _T \cdot \vec B_T) = + \theta B$ (more on this point in subsection 2c), with the present bound on $\theta$ one needs, at the beginning of the collision, a (non-commutative) photon with energy of about $1-2$  GeV at FCC and LHC to produce a  $e^+e^-$ pair at threshold. On the other hand, since $B$ is a function of time, the $e^+e^-$ invariant mass $m_{e^+e^-}$ depends on the  $\gamma$ production time, i.e.  $$ m_{e^+e^-} = E_\gamma \sqrt{2 B(t)} 10^{-4} GeV$$, where $B(t)$ and  $E_{\gamma}$ have been expressed in GeV unit. In fig.1 it is shown the time to reach the threshold for photons with different energies at LHC and FCC. The non-commutative contribution can only come from  hard parton scatterings at the beginning of the collision where the temperature and density effects of the quark gluon plasma are negligible.

\section{The experimental signatures}

\subsection{a. Signal to background ratio for large transverse momentum pairs}

Fig. 1 concerns the most favorable case and  since the magnetic field produced in the collisions is, event by event, orthogonal to the reaction plane \cite{dima2}, the non-commutative effects are maximized when the energy of the photon is in the reaction plane. Therefore the best signature for FCC is a increase of pair production with a very large total transverse energy in the reaction plane, between $100$ and $600$ GeV, to produce a detectable invariant mass $m_{e^+e^-}$ in the range  $100-400$ MeV  (see fig.1).

Let us try a rough estimate of this enhancement with respect to the background in peripheral collisions \cite{baur} given by the Compton-like quark-gluon scattering where a quasi real photon $\gamma^*$ decays in $ e^+ e^-$ and by the Dalitz decay, $\pi^0 \to \gamma e^+ e^-$, where $\pi^0$ are produced via gluon scatterings.

In the invariant mass range $0.2 < m_{e^+e^-} < 0.4 $ GeV one gets rid of the $\pi^0$ background, since the lepton pairs from Dalitz  decays of $\pi^0$ are below this mass range,  
and one stays below the $\rho,\omega,\phi \rightarrow e^+e^-$ threshold. Therefore the non-commutative contribution is essentially equal to the standard model calculation \cite{cerngroup} (for non-commutative photons there are only minor perturbative, non-commutative, correction to the differential cross section \cite{nc3}, the main effect being the modification of the theshold)) and one expects to double the number of lepton pairs at large transverse momentum in this mass range.

For lower invariant mass, i.e.  $ m_{e^+e^-} < 0.2$ GeV, the contribution due to  $\gamma^* \rightarrow e^+ e^-$ can be evaluated by the differential cross section for prompt real photon times the kinematical factor $\alpha \ln[M^2_{max}/(4 m_e^2)]/3\pi$ \cite{aurenche,cerngroup} where $M_{max}$ is the largest invariant mass allowed. The same method can be applied to evaluate the $e^+ e^-$  pairs produced by the non -commutative photon. Concerning the $\pi^0$ contribution, its differential cross section can be evaluated by taking into account the branching ratio
$B(\pi^0 \rightarrow \gamma e^+e^-/ \pi^0 \rightarrow \gamma  \gamma) \simeq 0.012$ \cite{bra} and that for $E^\gamma_T \ge 50$GeV the $\pi^0 \rightarrow \gamma \gamma$ differential cross section, for Pb-Pb collision at $\sqrt{s}= 5.5$ Tev, is essentially the same as the prompt real photon one \cite{cerngroup}.
Therefore the total differential cross section includes three contributions, the two standard effects (SB) just mentioned and the noncommutative one, all proportional to the prompt real photon cross section.
The ratio non-commutative signal to SB can be written as $N_{tot}/N_{SB} = 1+K_{NC}$ where $N$ is the number of $e^+e^-$ pairs and $K_{NC}$ 
is given by

\begin{eqnarray}
    K_{NC}= \frac{\frac{\alpha}{3 \pi} 
    \ln\left(\frac{2 E_\gamma ^2 (\vec \theta _T \cdot \vec B_T)}{4 m_e^2}\right)}{0.012 
    + \frac{\alpha}{3 \pi} \ln\left(\frac{2 E_\gamma ^2 (\vec \theta _T \cdot \vec B_T)}{4 m_e^2}\right)} \;, \label{estimate}
\end{eqnarray}
where $M^2_{max}$ has been identified with $2 E_\gamma ^2 |(\vec \theta _T \cdot \vec B_T)|$. 

At FCC , at the beginning of a gold-gold collision, $B=74.4$ GeV$^2$ and for $E_\gamma = p^\gamma_T = 100 $ GeV, the correction ( in the most favorable case) turns out 
$K_{NC} = 0.38$, i.e. a detectable enhancement of $e^+e^-$  pair production. 

However, the non-commutative correction has a time dependence, due to the fast decrease of the magnetic field, and it is less than $10 \%$ after 0.4 fm (see fig.2).

\begin{figure}
{{\epsfig{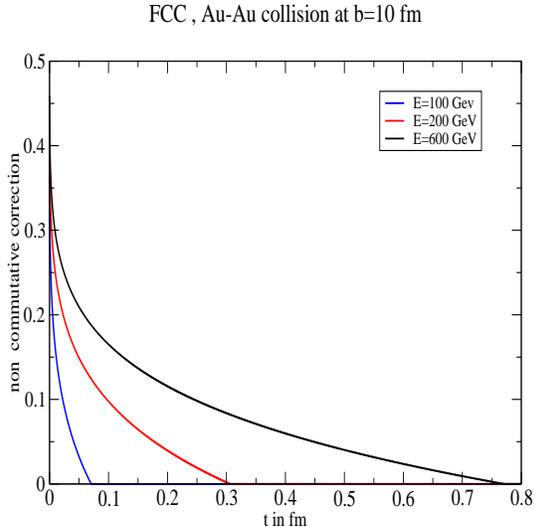}}
\caption{Non-commutative contribution as a function of time ( see text).
}}
\end{figure}

\subsection{b. Comparison with proton-proton}

In proton-proton (pp) collisions there is no magnetic field and therefore the ratio between the number of produced $e^+e-$ pairs in heavy ion and pp scatterings, with the same setting, i.e. large transverse energy in the reaction plane and small invariant mass, gives a signature of the non-commutative effects. Indeed , since one considers a very peripheral collision the nuclear modifications of the initial parton distributions are small, according to the local shadowing \cite{sandy,ramona} idea.

\subsection{c. Periodicity of the non-commutative signatures}

In the previous discussions the most favorable condition has been considered, which implies that the transverse component of the non-commutative vector parameter $\vec \theta$ with respect to the
photon momentum is parallel to the background field, i.e. it is orthogonal to the reaction plane. According to ref. \cite{volo}, this aspect would require a correlation analysis,  however there is a specific property of non-commutativity that could lead to a unique signature.

Defining the laboratory frame, say $(\hat{x},
\hat{y}, \hat{z})$, in such a way that the  reaction plane corresponds to the $\hat{x}-\hat{y}$ plane, the magnetic
field $B$ is produced in the $\hat z$ direction \cite{dima2}. The noncommutative effect is enhanced if the magnetic
field transverse to the direction of the momentum of the photon is maximum  and then one has to focus on the reaction plane with large transverse momentum $\gamma$.
 
On the other hand, the vector $\vec \theta$ is fixed in a non-rotating frame, denoted by $(\hat X, \hat Y, \hat Z)$, whereas
the component $\theta_z$ used in the most favorable condition is defined in the previously introduced frame.

Since this frame, at fixed reaction plane, rotates with the earth, this component changes in time with the periodicity that depends on
the earth's sidereal rotation frequency $\Omega$.

By following the choice in \cite{bound2,solito}, one can take
the $\hat Z$ direction of the non-rotating frame coincident with the rotation axis
of the earth and $\hat X $ and $\hat Y$ with specific fixed celestial equatorial coordinates.
Then, by indicating with $(\theta_X,\theta_Y,\theta_Z)$ the components of the noncommutative parameter
in the non-rotating frame, one gets the explicit time dependence of $\theta_z$ \cite{bound2}
\begin{equation}
\theta_z = (\sin \chi \cos \Omega t)\; \theta_X +
(\sin \chi \sin \Omega t) \; \theta_Y + \cos \chi \; \theta_Z
\end{equation}

where $\chi$ is the non-vanishing time-independent angle between the two axes $\hat Z$ and $\hat z$.
The oscillation in $\theta_z$ disappears in the peculiar case of $\vec \theta$ coincident with the
earth rotation axis (i.e. $\vec \theta=\theta_Z$ and therefore $\theta_z=\cos \chi \; |\vec \theta| $)
whereas it is maximal if $\vec \theta$ lies in the equatorial plane.

Apart from the unlikely case $\vec \theta = \theta_Z$,
Eq. (9) clearly shows the oscillating structure of the product $( \vec \theta _T \cdot \vec B_T)$
which appears in Eq. (7) and  which, for  photons considered above, reduces to the product $\theta_z B$.

Then the last signature of a nonzero $\vec \theta$ we are able to identify is a periodicity with frequency $\Omega$ in
the number of pairs produced at fixed reaction plane.

\section{Conclusions}

The FCC is a crucial step forward  not only to study new Physics beyond the standard model but also to test fundamental aspects as space-time non-commutativity and Lorentz violation.
In here, as an example, a possible enhamcement of $e^+e^-$ pair production due to non-commutative effect, catalyzed by the huge magnetic field produced at the beginning of a heavy ion collision at FCC,
is suggested in a particular kinematical setting: large total transverse momentum in the reaction plane and invariant mass in the rangge $100-400$ MeV. As discussed this requires a production of hard photons with energy $300-600$ GeV at the beginning of the collision. More generally, the magnetic background field could enhance the phenomenological effects of the violation of Lorentz invariance and/or of a minimal
lenght.

{\bf Acknowledgements} The author thanks the CERN Theory Department for the hospitality.


\begin{thebibliography}{99}

\bibitem{mich1} T.Gollin et al., Physics at a 100 TeV pp collider: beyond the standard model, arXiv: 1606.00947
\bibitem{mich2} M.Mangano et al., Physics at a 100 TeV pp collider: standard model processes, arXiv: 1607.01831
\bibitem{mich3} R.Contino et al. Editors, Physics at a 100 TeV pp collider: Higgs and EW symmetry breaking, arXiv: 1606.09408

\bibitem{armesto} A.Dainese et al., Heavy ions at the Future Circular Collider, arXiv: 1605.01389

\bibitem{nc1} M.~R.~Douglas, N.~A.~Nekrasov, Rev. Mod. Phys.  {\bf 73} (2001) 977;

\bibitem{nc2}R.~J.~Szabo, Phys. Rept. {\bf 378} (2003)  207;

\bibitem{nc3} I.~Hinchliffe, N.~Kersting and Y.~L.~Ma, Int. J. Mod. Phys. A {\bf 19} (2004) 179.



\bibitem{dima1} D.~E.~Kharzeev, Phys. Lett. B {\bf 633} (2006) 260; 

\bibitem{dima2} D.~E.~Kharzeev, L.~D.~McLerran and H.~J.~Warringa, Nucl. Phys. A {\bf 803} (2008) 227.

\bibitem{dima3} K.~Fukushima, D.~E.~Kharzeev and H.~J.~Warringa, Phys.Rev. {\bf D78} (2008) 074033


\bibitem{sko} V.~Skokov, A.~Y.~Illarionov and V.~Toneev, Int. J. Mod. Phys. A {\bf 24} (2009) 5925.

\bibitem{volo} S. A. Voloshin, {\it Local strong parity violation and new possibilities in experimental study of non-
perturbative QCD}, arXiv:1003.1127 [nucl-ex].

\bibitem{dima4} D.~E.~Kharzeev and A.~Zhitnitsky , Nucl. Phys.  {\bf A 797} (2007) 67.

\bibitem{dima5} K.~Fukushima, D.~E.~Kharzeev and H.~J.~Warringa, Nucl.Phys. {\bf A836} (2010) 311.

\bibitem{bernardini1}A. E. Bernardini and R. Da Rocha, Phys.Rev. D 75 (6),  (2007) 065014.

\bibitem{bernardini2} A. E. Bernardini and O. Bertolami, Phys. Rev. A88,  (2013) 012101.

\bibitem{bernardini3} C. Bastos, A. E. Bernardini, O. Bertolami, N. Dias and J. Prata, Phys. Rev. A89, (2014) 042112 ; Phys. Rev. D90,
(2014) 045023 ; Phys. Rev. D91,  (2015) 065036; Phys. Rev. D93,  (2016) 104055.

\bibitem{roman} Z.~Guralnik, R.~Jackiw, S.~Y.~Pi, A.~P.~Polychronakos, Phys. Lett.  B {\bf 517} (2001) 450.

\bibitem{noi1} P.~Castorina, A.~Iorio and D.~Zappal\`a,  Phys. Rev. D {\bf 69} (2004) 065008.

\bibitem{noi2} P.~Castorina, A.~Iorio and D.~Zappal\`a, Europhys. Lett. {\bf 69} (2005) 912.

\bibitem{noi3} P.~Castorina, and D.~Zappal\`a, Europhys. Lett.  {\bf 64} (2003) 641;

\bibitem{noi4} P.~Castorina, A.~Iorio, D.~Zappal\`a, Nucl. Phys. Proc. Suppl. {\bf 136} (2004) 333.

\bibitem{sw} N.~Seiberg and E.~Witten, J. High Energy Phys.  {\bf 9909} (1999) 032.

\bibitem{solito} S.~M.~Carroll, J.~A.~Harvey, V.~A.~Kostelecky, C.~D.~Lane, and T.~Okamoto,  Phys. Rev. Lett. {\bf 87} (2001) 141601.

\bibitem{orfeo1}  O.~Bertolami et al., Phys. Rev. D {\bf 72} (2005) 025010.

\bibitem{orfeo2}  O.~Bertolami et al., Mod.Phys.Lett. A {\bf 21} (2006) 795.

\bibitem{noi5} P.~Castorina, A.~Iorio, D.~Zappal\`a, Eur.Phys.J. {\bf C71} (2011) 1653.

\bibitem{baur} G. Baur, K. Hencken, D. Trautmann, Phys. Rept. {\bf 453} (2007) 1.

\bibitem{aurenche} P.~Aurenche, B.~Baier, M.~Fontannaz, Phys. Lett. B {\bf 209} (1988) 375.

\bibitem{cerngroup} F.~Arleo et al., {\it Writeup of the working group Photon Physics for the CERN Yellow Report on Hard Probes in Heavy Ion
Collisions at the LHC}, arXiv:hep-ph/0311131

\bibitem{bra} A. Beddall,  Eur. Phys. J.  {\bf C54} (2008) 365.

\bibitem{sandy} P.Castorina and A.Donnachie, Z.Phys. {\bf C49} (1991) 481.

\bibitem{ramona} S. R. Klein and R. Vogt, Phys. Rev. Lett. {\bf 91} (2003) 142301.

\bibitem{bound2} V.~A.~Kostelecky and  C.~Lane, Phys. Rev. D {\bf 60} (1999) 116010.




\end{thebibliography}
\end{document}